\documentclass[twocolumn,superscriptaddress,amsmath,amssymb,showpacs,prl]{revtex4-1}

\usepackage{graphicx}
\usepackage{color}
\renewcommand{\phi}{\varphi}

\begin{document}

\title{Deep-learning-based prediction of the tetragonal→cubic transition in davemaoite}

\author{Fulun Wu}
	\affiliation{Department of Physics, Xiamen University, Xiamen 361005, China}
\author{Yang Sun}
	\affiliation{Department of Physics, Xiamen University, Xiamen 361005, China}
	\affiliation{Department of Applied Physics and Applied Mathematics, Columbia University, New York, NY 10027, USA}
\author{Tianqi Wan}
	\affiliation{Department of Applied Physics and Applied Mathematics, Columbia University, New York, NY 10027, USA}
\author{Shunqing Wu}
	\affiliation{Department of Physics, Xiamen University, Xiamen 361005, China}
\author{Renata M. Wentzcovitch}
	\affiliation{Department of Applied Physics and Applied Mathematics, Columbia University, New York, NY 10027, USA}
	\affiliation{Department of Earth and Environmental Sciences, Columbia University, New York, NY 10027, USA}
	\affiliation{Lamont–Doherty Earth Observatory, Columbia University, Palisades, NY 10964, USA}

\date{Jan. 26, 2024}

\begin{abstract}

Davemaoite, i.e., $CaSiO_3$ perovskite (CaPv), is the third most abundant phase in the lower mantle and exhibits a tetragonal-cubic phase transition at high pressures and temperatures. The phase boundary in CaPv has recently been proposed to be close to the cold slab adiabatic and cause mid-mantle seismic wave speed anomalies (Thomson et al., Nature 572, 643, 2019). In this study, we utilized accurate deep-learning-based simulations and thermodynamic integration techniques to compute free energies at temperatures ranging from 300 to 3,000 K and pressures up to 130 GPa. Our results indicate that CaPv exhibits a single cubic phase throughout lower-mantle conditions. This suggests that the phase diagram proposed by Thomson et al. requires revision, and mid-mantle seismic anomalies are likely attributable to other mechanisms.

\end{abstract}

\maketitle
\textit{Introduction}.\textemdash $CaSiO_3$ perovskite (CaPv) is the third most abundant phase in a pyrolitic lower mantle and accounts for 24-29 vol$\%$ and 5-10 vol$\%$ of basaltic and pyrolitic assemblages, respectively  \cite{1,2}. Thermodynamic and thermoelastic properties of CaPv phases are crucial for a better understanding of Earth's interior, particularly for understanding potential compositional heterogeneities in the mantle and the $D''$ region where basaltic compositions are expected to be abundant  \cite{3,4,5,6}. Several experiments and theoretical studies have been carried out to unravel CaPv's structure and thermodynamic properties  \cite{7,8,9,10,11,12,13,14,15}. With increasing temperature, CaPv undergoes a tetragonal ($I4/mcm$) to cubic ($Pm\overline{3}m$) phase transition. Recent works have confirmed that cubic CaPv is stabilized by its strong anharmonic fluctuations at high temperatures  \cite{13,14,15}. Because tetragonal and cubic CaPv have rather different sound velocities  \cite{3,16}, their stability field under mantle conditions can be crucial to understanding velocity heterogeneities in the mantle if this transition occurs at mantle conditions.

So far, the tetragonal-to-cubic phase boundary of CaPv under mantle pressures remains controversial. Stixrude et al. first predicted that the tetragonal-to-cubic CaPv transition occurs at temperatures above 2200 K and pressures of 80 GPa  \cite{10}. However, Kurashina et al. conducted an experimental study and observed the phase transition at a much lower temperature of approximately 580 K and 50 GPa  \cite{17}. It is important to note that the temperature measurements in Kurashina et al. had a significant uncertainty. Follow-up experiments conducted by Ono et al.  \cite{18} and Murakami et al.  \cite{19} using laser-heated DAC with synchrotron X-rays diffraction suggested the transition of CaPv to a cubic structure above 2000 K, while it maintained a tetragonal structure at room temperature. However, these studies did not provide precise phase boundaries at lower mantle conditions. In 2007, Komabayashi et al. performed X-ray diffraction experiments and determined that the transition temperature of CaPv ranged from 490 to 580 K at pressures between 27 and 72 GPa  \cite{8}, consistent with the findings of Ono et al. and Kurashina et al.. However, a study conducted in 2009 presented conflicting results, suggesting that CaPv remains a non-cubic phase up to 18 GPa and 1600 K  \cite{9}.Theoretical studies conducted by Adams using \textit{ab initio} molecular dynamics in a metrically cubic supercell found the transition temperature to cubic CaPv to be between 1000 and 2000 K  \cite{20}. Similar conclusions were reached by Stixrude et al. using Landau mean-field potentials  \cite{21}. However, Li et al. conducted AIMD and found that the hydrostatically compressed structure remained tetragonal up to 4000 K  \cite{12}. Sun et al. demonstrated that cubic CaPv is stabilized at approximately 600 K and 26 GPa  .\cite{13}. Overall, the experimental characterization of the phase boundary is challenging because cubic CaPv is not quenchable to ambient conditions  \cite{8,9,13}. \textit{Ab initio} calculations of the phase boundary of CaPv are also highly nontrivial due to strong anharmonicity in this system \cite{10,13,16}. Free energy approaches, such as thermodynamic integration (TI)  \cite{22} based on purely \textit{ab initio} molecular dynamics (AIMD) can be used but is computationally intensive, thus only practical for small systems.

In 2019, Thomson et al. observed the tetragonal-to-cubic phase transition in CaPv at ~12 GPa through in situ diffraction and identified the reduction of the seismic velocity along with the phase transition  \cite{16}. To understand the geophysical consequence of this transition, they combined the experimental measurement and \textit{ab initio} calculations to determine the phase boundary of CaPv at lower mantle pressures. The slope of the CaPv phase boundary was estimated by \textit{ab initio} free energy calculations as dT/dP $\sim$15 K/GPa. Using the calculated dT/dP, they extrapolated the experimental data from 12 GPa to the entire lower mantle pressures and found the CaPv undergoes a cubic-tetragonal transition in cold slab assemblages at pressures greater than 90 GPa. Their phase boundary thus was proposed to explain the seismic anomalies in the mantle, such as the mid-mantle reflectors  \cite{16}. However, their theoretical slope of the phase boundary was obtained via an approximated thermodynamic integration method and the data deviated significantly from Komabayashi et al.'s experimental value of ~1.1±1.3 K/GPa  \cite{8}. 

Considering the lack of consensus between these studies, it is evident that further investigations are required to gain a deeper understanding of the phase transition in CaPv, though purely \textit{ab initio}  calculations remain challenging. Recently, interatomic potentials developed with machine learning techniques such as Neural Network Potential (NNP)  \cite{23}, Gaussian Approximation Potential (GAP)  \cite{24}, on-the-fly Machine Learning Force Field (MLFF)  \cite{25}, and Deep Potential for Molecular Dynamics (DPMD)  \cite{26} have significantly extended the timescale and length scale of \textit{ab initio}  simulation. These algorithms can incorporate large amounts of computational or experimental data to construct direct mappings from atomic structures to forces and energies, thus saving significant amounts of computational time required for \textit{ab initio}  calculations  \cite{23,24,27,28,29,30,31,32,33,34}. Particularly, the DP method has recently shown its ability to capture higher-order anharmonic force constant  \cite{35}, which can be useful in studying the high-temperature properties of strongly anharmonic materials. By combining with molecular dynamics, DP has proven to be effective in the study of several systems under high-PT conditions, e.g., $MgSiO_3$  \cite{35,36,37,38}, $FeSiO$ melts  \cite{39,40}, $\delta-AlOOH$  \cite{41}, dense lithium  \cite{42}, high-pressure phases of tin  \cite{43}, solid and fluid $H_2O$  \cite{44} etc.

In this work, we use the DP method to study phase transitions in CaPv at temperatures ranging from 300-3000K and pressures up to 130 GPa. We develop DP potentials with an active learning method and demonstrate its accuracy in predicting energies, forces, and vibrational properties at various PT conditions. By combining the DP and the TI methods, we compute the phase boundary between the tetragonal and cubic phases of CaPv throughout the lower mantle's pressure range and discuss its implication on the seismic anomaly in the lower mantle. 

\textit{Methods}.\textemdash \textit{Ab initio} calculations were performed using the Vienna Ab-initio Simulation Package (VASP)  \cite{45,46}, which implements the projector-augmented wave (PAW)  \cite{47,48} methodology of density functional theory. Functionals with the local density approximations (LDA)  \cite{49} and the generalized gradient approximation (GGA)  \cite{48} were studied. PAW potentials with valence electronic configurations $3s^23p^6$, $3s^23p^2$, and $2s^22p^4$ for Ca, Si, and O were used. A plane wave energy cutoff of 550 eV and a 2×2×2 k-point grid for the 40-atom supercell were used, which was shown to be sufficient to achieve convergence  \cite{15}. Two independent DP models, using LDA and GGA datasets, were developed and are referred to as DP-LDA and DP-GGA, respectively. The smooth edition descriptor $se\_e2\_a$ proposed by Zhang et al.  \cite{26,50} was employed to train these models (see training details in Supporting Information Text S1).

The harmonic phonon spectrum was calculated using the finite displacement supercell approach implemented in the Phonopy  \cite{51}. Anharmonic effects were addressed by the phonon quasiparticle method  \cite{52}. Temperature-dependent anharmonic phonon dispersions were computed by projecting the atomic velocities onto the harmonic phonon eigenvectors of the temperature-dependent equilibrium structure. In this method, the phonon frequencies and corresponding linewidths can be extracted directly from the MD trajectories, as described by  \cite{13}. The DP-based molecular dynamics (DPMD) simulations were used to obtain the mode-projected velocity auto-correlation functions (VAF)  \cite{53}.

To calculate the free energies of the $I4/mcm$ and $Pm\overline{3}m$ phases of CaPv, we used a state-of-the-art TI method developed by  \cite{54}. It provides an efficient method to sample the full TI path from the reference state and target state. Using DPMD simulations, we numerically integrated the work from an initial equilibrium state to the final equilibrium state of interest. To obtain the Gibbs free energy,

\begin{equation}
	G_{CaPv}=F_{CaPv}+PV
\end{equation}

\noindent we compute the Helmholtz free energy $F_{CaPv}$ by

\begin{equation}
	F_{CaPv}\left(N,V,T\right)=F_E\left(N,V,T\right)+\Delta F
\end{equation}

\noindent where $F_E\left(N,V,T\right)$ represents the Einstein crystal reference. The Hamiltonian of this harmonic reference system can be written as

\begin{equation}
	H_E = \sum_{i=1}^{N}\left[\frac{{p}_i^2}{2m}+\frac{1}{2}m\omega^2\left({r}_i-{r}_i^0\right)^2\right]
\end{equation}

\noindent where $\omega$ is the oscillator frequency and $r_i^0$ is the equilibrium position of particle i. Its Helmholtz free energy is given by

\begin{equation}
	F_E\left(N,V,T\right)=3k_BT\ln{\left(\frac{\hbar\omega}{k_BT}\right)}
\end{equation}

In order to closely match the characteristic vibrational spectrum of the CaPv system, the value of $\omega$ was obtained through high-temperature molecular dynamics simulations. The mean-squared displacement $\langle (\Delta r)^2 \rangle$  of atoms in the CaPv system was used to determine $\omega$ based on the equipartition theorem as $m\omega^2=\frac{3k_BT}{\langle(\Delta r)^2\rangle}$  \cite{54}. This approach differs from the method employed in  [16], where $\omega$ was obtained using harmonic force constants with the artificial elimination of imaginary phonon modes. Since our molecular dynamics simulations already incorporate the effects of anharmonicity, the determination of $\omega$ using the mean square displacement (MSD) method introduced by  \cite{54} provides a more realistic reference state. The TI method  \cite{54} provides a continuous path from the Einstein crystal to CaPv. The forward and backward integration processes were employed to enhance the sampling of the transition pathway. The free energy difference $\Delta F$ was computed by

\begin{equation}
	\Delta F=\frac{1}{2}\left(W_{ref-final}^{irr}-W_{final-ref}^{irr}\right)
\end{equation}

\noindent where $W^{irr}$ is the irreversible work to transform from the Einstein crystal to the CaPv, which is noted as

\begin{equation}
	W_{ref-final}^{irr}=\int_{0}^{1}{\left.\ \left\langle\ \frac{\partial H\left(\lambda\right)}{\partial\lambda}\ \right.\ \right\rangle_\lambda\ \ d\lambda}
\end{equation}

\noindent $\lambda$ is the coupling parameter between Einstein crystal and the CaPv. A supercell of 2560 atoms was used to compute the $\Delta F$ for both tetragonal and cubic phases of CaPv. We noted that at the transition temperatures, the tetragonal phase is metastable and remains the c/a unchanged in the MD simulations.

\textit{Phonon dispersion and equation of states}.\textemdash We collected 3420 unique configurations from the \textit{ab initio}  data exploration in 300-3000 K and 20-130 GPa with 40-atom supercell. We split the data set into 80$\%$ for training and 20$\%$ for validation. To evaluate the accuracy of the DP, the root mean square errors (RMSEs) of energies, atomic forces, and pressures were compared to DFT calculations for the CaPv system. The results, as shown in Figure S1(a-c), suggest excellent agreement between DP and DFT calculations, with RMSEs of about 2.0 meV/atom for energies, 0.052 eV/Å for forces and 0.2 GPa for pressures. Similar accuracy was also achieved for DP-GGA, as shown in Figure S1(d-f). The RMSEs of the training and validating sets are similar, suggesting the neural network was not overfitting. Furthermore, the wide energy distribution displayed in Figure S1(a) and (d) indicates the complexity of the configuration spaces involved in exploring the potential energy surface.

The sensitivity of the phonon dispersion relations to the accuracy of atomic interactions stems from their dependence on the quality of the force constant matrix (dynamical or Hessian matrix). The DP potentials were then applied to determine the interatomic force constants and dynamical matrices via the finite displacement method  \cite{51}. Figure 1 demonstrates that the DP potential accurately reproduces the harmonic phonon spectra for both tetragonal and cubic CaPv phases. It is noteworthy that the DP method accurately reproduces the soft phonon modes at \textbf{M}(1/2, 1/2, 0) and \textbf{R}(1/2,1/2,1/2) in cubic CaPv at 0 K, as shown in Figure 1(a). Similar results are also found for DP-PBE potentials, shown in Figure S2. While soft and unstable modes are present at 0 K in the cubic phase, they can be stabilized at higher temperatures by anharmonic effects  \cite{13}. The temperature-dependent phonon dispersions at 1500K, 2500K, and 3500K were then calculated using the phonon quasiparticle method  [52] with the DP model. As shown in Figure 1(b), while most phonon modes showed a weak temperature dependence, significant frequency renormalization was observed at \textbf{M}(1/2, 1/2, 0) and \textbf{R}(1/2,1/2,1/2) as the temperature increased, due to strong lattice anharmonicity. In comparison to the harmonic phonon spectrum for cubic CaPv in Figure 1(a), the imaginary frequencies at M and R were lifted to positive frequencies, indicating the dynamic stability of the structure. This reproduces the LDA results in Zhang et al.  \cite{15}. Similar frequency renormalization is also found with the DP-GGA. The accurate prediction of the anharmonic phonon dispersions demonstrates the ability of our DP to describe both the harmonic lattice dynamics and high-temperature effect, which is crucial for accurately capturing the phase transitions.

\begin{figure}
\includegraphics[width=0.49\textwidth]{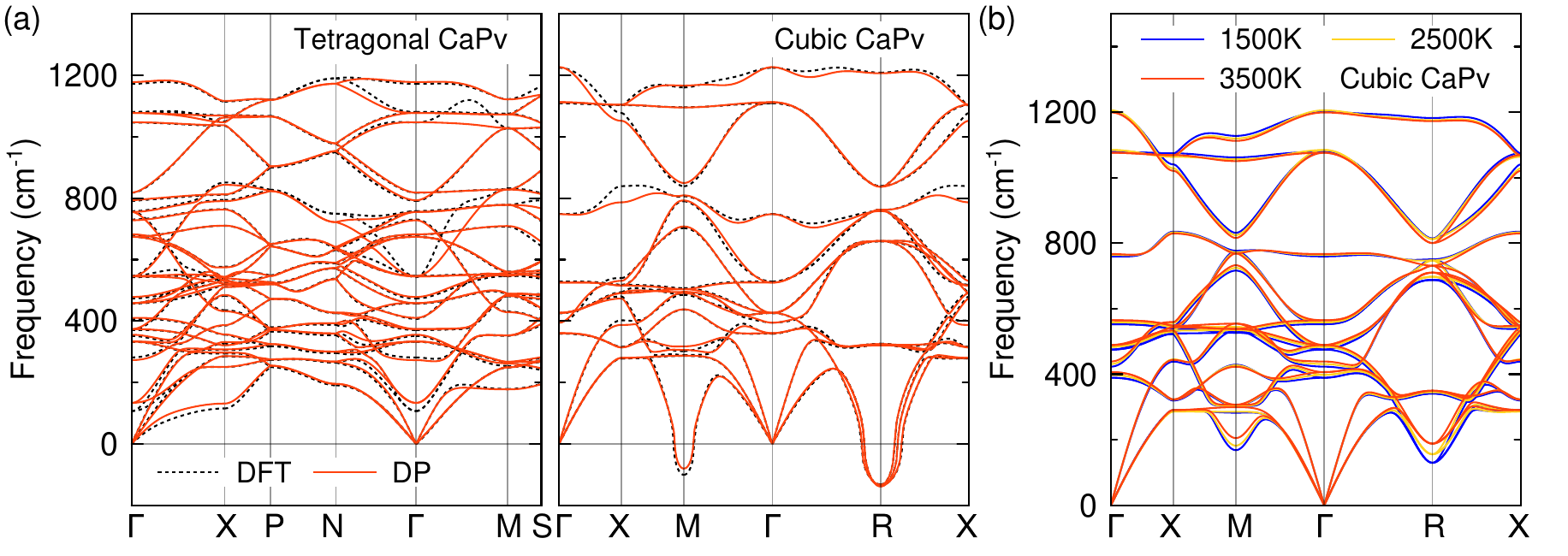}
\caption{(a) Harmonic phonon dispersions of tetragonal and cubic CaPv at 70GPa and 0K along high-symmetry lines in the Brillouin zone by DP and DFT with LDA functional. (c) Anharmonic phonon dispersions of cubic CaPv at temperatures of 1500K, 2500K, and 3500K by DP.}
\end{figure}

We further compare the equation of state (EOS) of CaPv with previous results to validate DP potential across a wide range of P-T conditions. Figure 2(a) and (b) show that the EOSs of cubic CaPv from our DP-LDA and DP-GGA agree well with previous DFT calculations  \cite{12,55,56} at 300 K and 2000 K. The experimental EOS data  \cite{3,7,16,57,58,59,60,61} are mostly between the LDA and GGA results. It is expected because the GGA functional tends to overestimate the volume, while the LDA approach produces underestimated volumes. We note that the difference in EOS between the tetragonal and cubic phases of CaPv is very small, as shown in Supplementary Figure S3.

\begin{figure}
\includegraphics[width=0.49\textwidth]{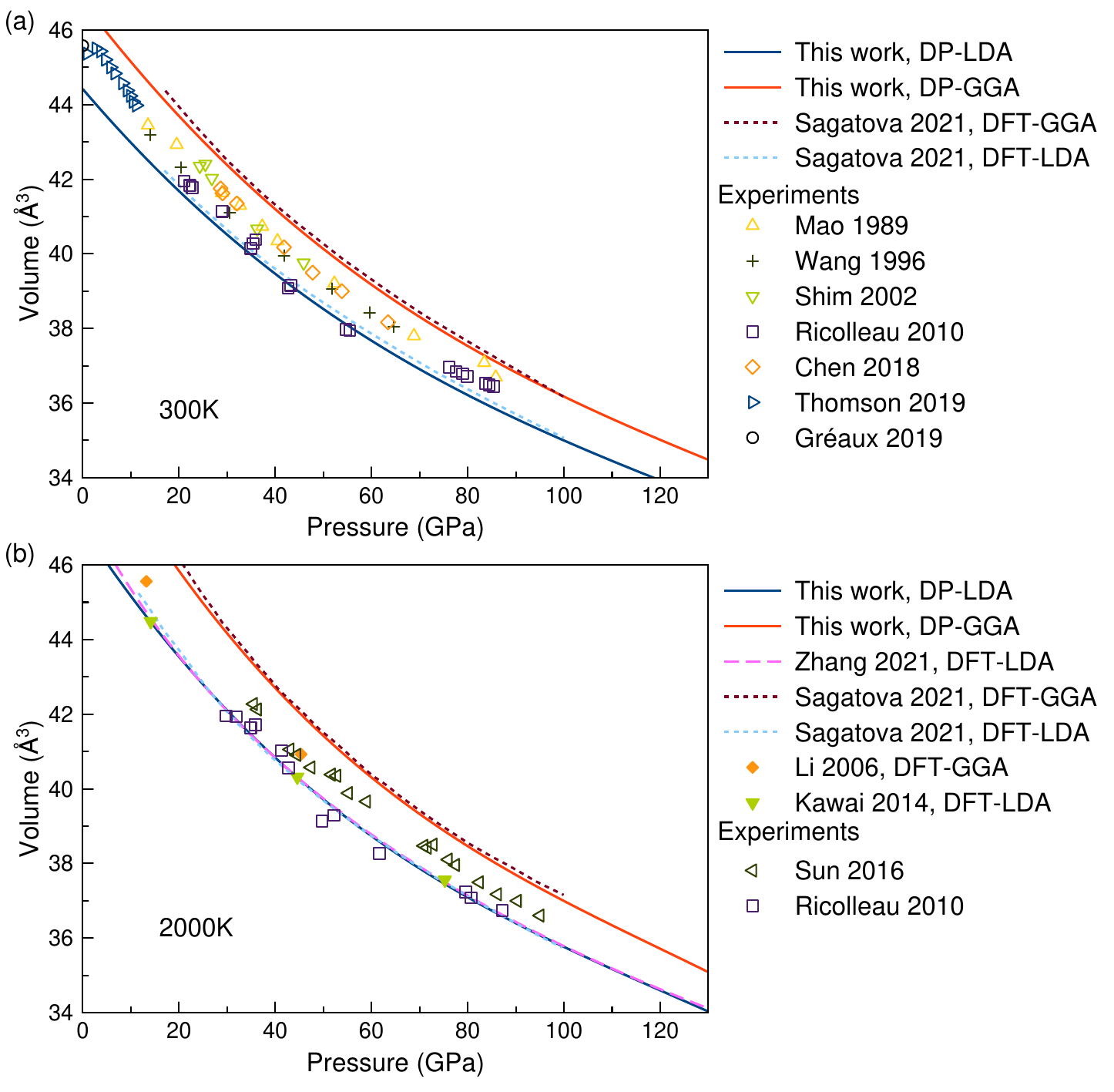}
\caption{Equation of state of cubic CaPv at (a) 300 K and (b) 2000 K. Solid curves are obtained using DP, and dashed lines are the DFT results \cite{12,15,55,56}, and experimental data are taken from Refs. \cite{3,7,16,57,58,59,60,61}.}
\end{figure}

\textit{Phase transition of CaPv}.\textemdash To describe the relative stability between the tetragonal and cubic phases of CaPv, we compute their Gibbs free energy difference $\Delta G\left(T,p\right)=G_{tet}\left(T,P\right)-G_{cub}\left(T,P\right)$ at different P-T conditions. A negative $\Delta G$ indicates that the tetragonal phase is more favorable and vice versa. We employ the TI method to compute the $G_{tet}$ and $G_{cub}$ for tetragonal and cubic phases, respectively. The DP-based TI-MD was performed at different temperatures and pressures for the two phases. The energy differences between the CaPv and Einstein crystal reference, denoted as $\Delta U(\lambda)=\langle U_{CaPv}-U_{ref}\rangle$, were obtained from TI-MD. As shown in Figure 3, the transition paths from the Einstein crystal ($\lambda=0$) to the CaPv ($\lambda=1$) were smooth for both tetragonal and cubic phases. The highly nonlinear transition path suggests a significant difference between the two systems. Therefore, one must sample this transition path with a continuously changing $\lambda$, which is achieved by the present TI method. Previous studies suggested the size effect can be significant in TI  \cite{62}. To examine this effect, we conducted the TI simulations with different numbers of atoms for both phases in Figure 4(a). It shows a simulation size larger than 2,000 atoms can converge the free energy calculation within 0.1 meV/atom.

\begin{figure}
\includegraphics[width=0.49\textwidth]{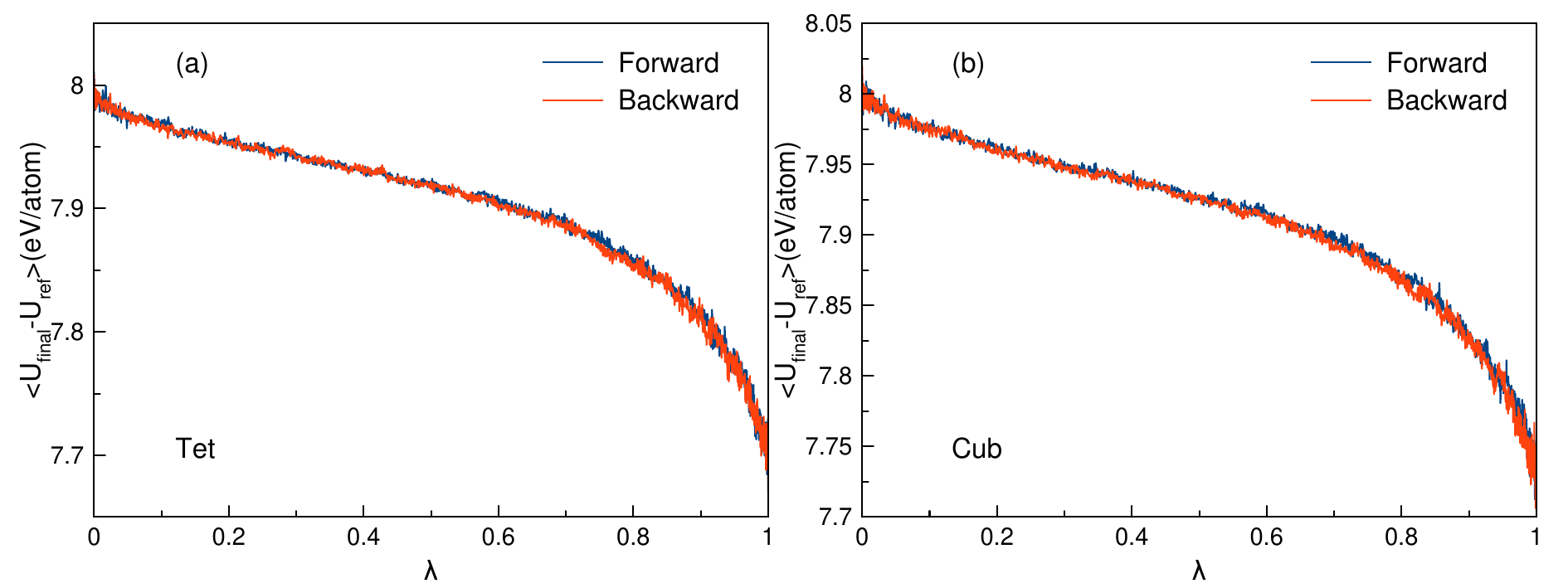}
\caption{TI calculations at 50 GPa and 1200K. (a-b) show the averaged energy differences between the final system ($\lambda=1$) and reference system ($\lambda=0$) in TI-MD simulations for tetragonal and cubic CaPv, respectively.}
\end{figure}

We show $\Delta G(T)$ with DP-LDA at 50 GPa in Figure 4(b). More $\Delta G(T)$ results at various pressure are shown in the Supplementary Figure S4. The tetragonal-cubic phase transition temperatures $T_c$ were obtained from the conditions where $\Delta G(T_c)$ = 0. For DP-LDA, the phase boundary exists around 415 K at 25 GPa, 510 K at 50 GPa, 670 K at 75 GPa, 930 K at 100 GPa, and 1240 K at 125 GPa. We also computed $\Delta G(T)$ with DP-GGA in Figure S5. For DP-GGA, the phase boundary is around 270 K at 25 GPa, 365 K at 50 GPa, 515K at 75 GPa, 680 K at 100 GPa, and 875 K at 125 GPa. It is worth mentioning that even if the most advanced massively parallel computers are employed, carrying out these simulations using full \textit{ab initio}  calculations is prohibitively time-consuming.

\begin{figure}
\includegraphics[width=0.49\textwidth]{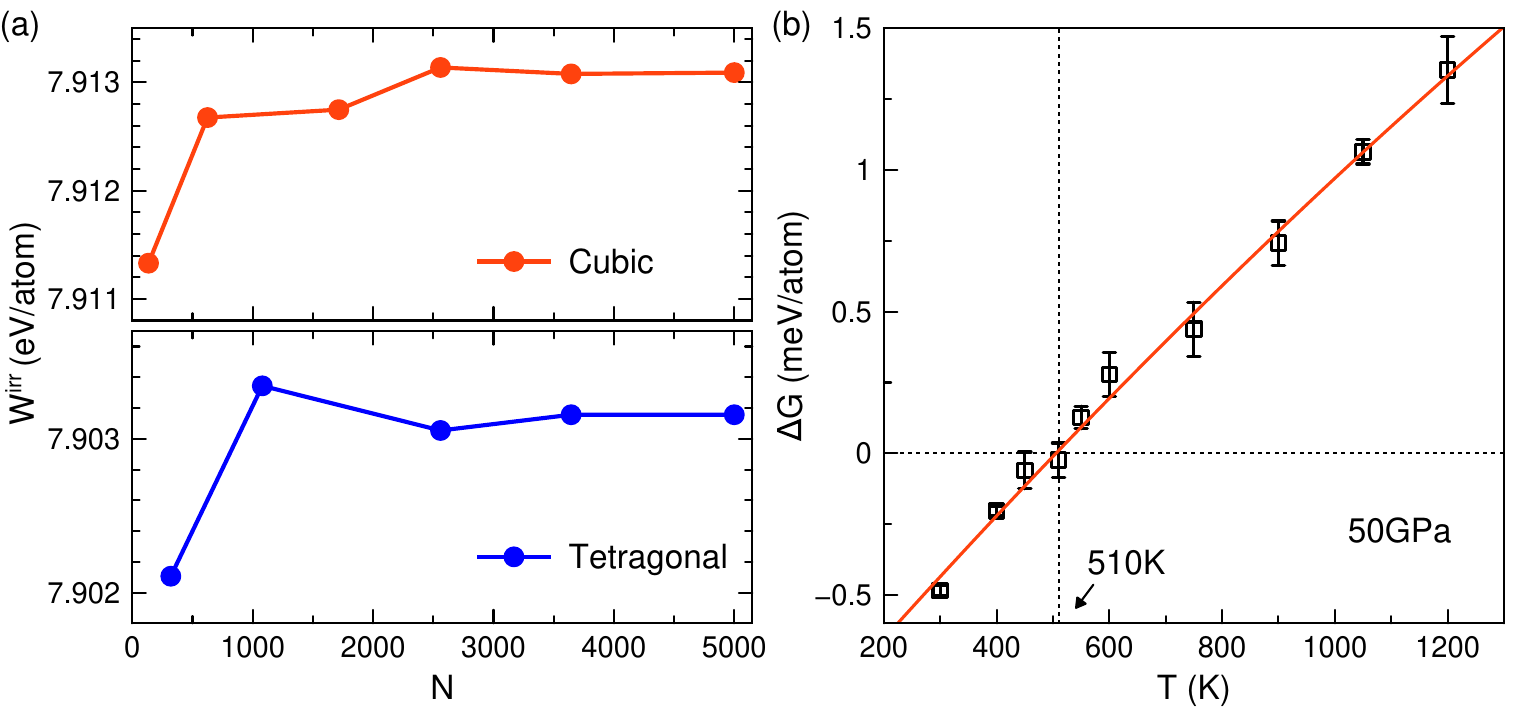}
\caption{(a) Size effect on the irreversible work with DP-based TI simulations. The circles are $W^{irr}$ measured from the TI with different numbers of atoms, N, in the simulations at 50 GPa and 1200K. (b) The Gibbs free energy difference between the tetragonal and cubic phase for the CaPv at 50 GPa (See Figure S2 for more pressures). Negative energy difference indicates that the tetragonal phase is more favorable, and vice versa.}
\end{figure}

Figure 5 shows the tetragonal-cubic phase boundary of CaPv obtained from our DP calculations. The DP-LDA results in a higher transition temperature than the DP-GGA by 200K to 500K. The differences increase with increasing pressure. Nevertheless, both phase boundaries from LDA and GGA calculations are significantly lower than the earlier calculations from Stixrude et al.  \cite{21} and Thomson et al.  \cite{16} by more than 1000 K under lower mantle pressures. The high computational cost of full \textit{ab initio} calculations has forced these studies to rely on approximated free energy calculations. Stixrude et al.  \cite{21} used a Landau potential that relies on a few empirical parameters to fit the \textit{ab initio}  data. The accuracy of this potential may not be sufficient to describe the small free energy difference between the cubic and tetragonal phases. Thomson et al.  \cite{16} used a perturbation method to apply a first-order approximation on the TI as the energy difference between the reference state and CaPv, as $F-F_0\cong\left\langle U-U_0\right\rangle_0+\frac{1}{2k_BT}\ \left\langle\left[U-U_0-\left\langle U-U_0\right\rangle_0\right]^2\right\rangle_0$. The accuracy of this equation highly depends on the linearity of the TI path. However, as shown in Figure 3, the TI path from the reference state to CaPv is highly nonlinear, so such an approximation is inaccurate for the CaPv system. Sagatova's DFT result  \cite{56} is similar to ours, where a self-consistent phonon algorithm (SCPH)  \cite{63,64} was used to perform anharmonic phonon calculations.

\begin{figure}
\includegraphics[width=0.49\textwidth]{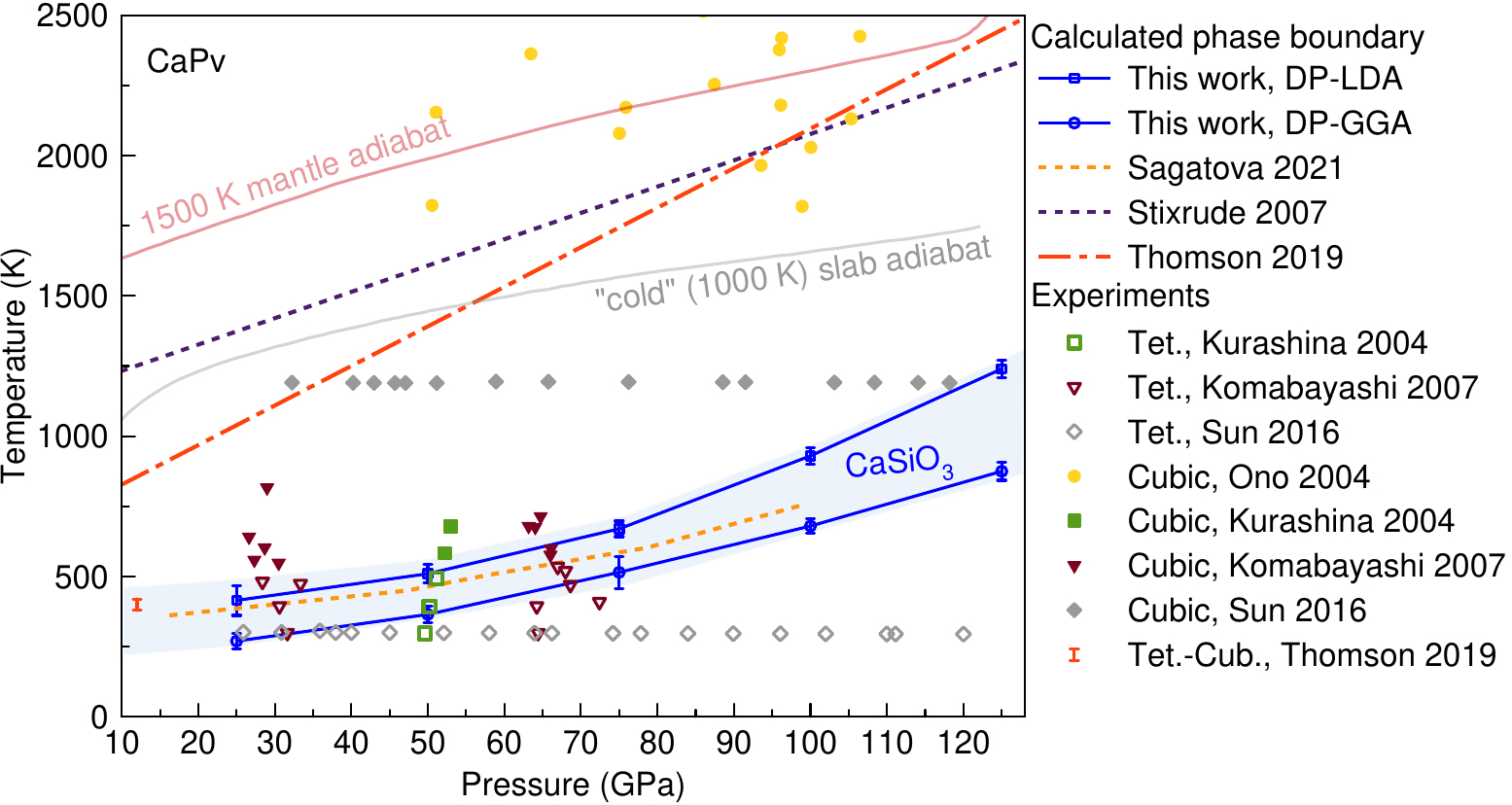}
\caption{Phase diagram of Ca-perovskite across the mantle pressure range. Previous data includes calculations by  \cite{16,21,56} and experiments by  \cite{8,16,17,18,57}. The shaded blue area indicates the uncertainty due to the DFT functional. The brown and gray solid lines depict 1,500 K mantle adiabat and 1,000 K “cold” slab temperature profiles, respectively. }
\end{figure}

The difference between LDA and GGA functionals can be treated as the computational uncertainty of the present calculated phase boundary, which provides upper and lower bounds. We consider this region as the confident interval of our calculations due to DFT functionals (blue region in Figure 5). Compared to experimental data, our phase boundary is consistent with all previous measurements  \cite{8,17,18}, including Thomson et al.'s experimental results at 12 GPa  \cite{16}. The dT/dP slope is $\sim$ 7 K/GPa, significantly smaller than the calculated results in  \cite{16}. 

Our tetragonal-to-cubic boundaries of CaPv lie well below the cold slab geotherm  \cite{7}, where CaPv is supposed to be abundant, or any adiabatic mantle geotherm, suggesting that CaPv should be cubic in all lower mantle environments. Therefore, the tetragonal to cubic transition in CaPv cannot explain the seismic wave speed anomalies in the lower mantle, contrary to the claim of  \cite{16}. The mid-mantle reflectors or scatters may be caused by other mechanisms, e.g. the post-stishovite phase transition in pure or aluminous-hydrous silica  \cite{6,65} in stagnant subducted slabs  \cite{66}. Recently, Ko et al. suggested that the dissolution behavior of CaPv in bridgmanite can lead to its disappearance as a separate phase at depths greater than 1,800km, a new insight into the deep mantle mineralogy  \cite{67}. Irrespective of this dissolution behavior, the problem addressed here is relevant for addressing the origin of seismic scatterers above 1800 km depth.

\textit{Conclusion}.\textemdash In summary, we have implemented DP models for $CaSiO_3$ perovskite using both LDA and GGA functionals. The simulations with the DP model reproduce the harmonic and anharmonic phonon dispersions from DFT calculations. The comparison with experiments suggests volume vs. pressure in EOS is underestimated by the LDA and overestimated by the GGA functionals. Overall, the machine learning approach can describe the strongly anharmonic properties of CaPv with \textit{ab initio} accuracy. Based on the DP model, we employ the TI method to fully sample the transition path from the reference state of a harmonic crystal to the CaPv phases. The transition path shows strong nonlinearity w.r.t. the integration parameter $\lambda$ which suggests TI in this system cannot be approximated as the energy difference between the initial and final states. Therefore, the previously calculated phase boundary in  \cite{16} is questionable. With the present DP model and TI method, we provide accurate free energy difference between the cubic and tetragonal phases in a large pressure and temperature range close to lower mantle conditions. We determined the tetragonal-to-cubic transition temperatures vs. pressure in CaPv and computed the thermodynamic phase boundary. We can satisfactorily explain previous experimental measurements. Based on our phase boundary, the CaPv are cubic at lower mantle conditions. It suggests the observed mid-mantle seismic scatters cannot be explained by the tetragonal-to-cubic phase transition in CaPv and is likely attributable to other mechanisms.

\begin{acknowledgments}
\textit{Acknowledgments}.\textemdash We are grateful to Rodrigo Freitas for the help on the thermodynamic integration method. Work at Xiamen University was supported by NSFC (Grants No. 12374015 and No. 42374108). Work at Columbia University was supported by NSF (Grants No. EAR-2000850 and No. EAR-1918126). Work at Iowa State University was supported by NSF Grant No. EAR-1918134. Shaorong Fang and Tianfu Wu from the Information and Network Center of Xiamen University are acknowledged for their help with the Graphics Processing Unit (GPU) computing.
\end{acknowledgments}

\bibliographystyle{apsrev4-1}

\begin{thebibliography}{46}%
\bibitem{1}	W. F. McDonough and S. -s. Sun, The Composition of the Earth, Chem. Geol. 120, 223 (1995).
\bibitem{2}	J. J. Valencia-Cardona, G. Shukla, Z. Wu, C. Houser, D. A. Yuen, and R. M. Wentzcovitch, Influence of the Iron Spin Crossover in Ferropericlase on the Lower Mantle Geotherm, Geophys. Res. Lett. 44, 4863 (2017).
\bibitem{3}	S. Gréaux, T. Irifune, Y. Higo, Y. Tange, T. Arimoto, Z. Liu, and A. Yamada, Sound Velocity of CaSiO3 Perovskite Suggests the Presence of Basaltic Crust in the Earth’s Lower Mantle, Nature 565, 7738 (2019).
\bibitem{4}	K. Hirose, N. Takafuji, N. Sata, and Y. Ohishi, Phase Transition and Density of Subducted MORB Crust in the Lower Mantle, Earth Planet. Sci. Lett. 237, 239 (2005).
\bibitem{5}	S. E. Kesson, J. D. Fitz Gerald, and J. M. G. Shelley, Mineral Chemistry and Density of Subducted Basaltic Crust at Lower-Mantle Pressures, Nature 372, 6508 (1994).
\bibitem{6}	W. Wang, Y. Xu, D. Sun, S. Ni, R. Wentzcovitch, and Z. Wu, Velocity and Density Characteristics of Subducted Oceanic Crust and the Origin of Lower-Mantle Heterogeneities, Nat. Commun. 2020 111 11, 1 (2020).
\bibitem{7}	S.-H. Shim, R. Jeanloz, and T. S. Duffy, Tetragonal Structure of CaSiO3 Perovskite above 20 GPa: STRUCTURE OF CaSiO3 PEROVSKITE, Geophys. Res. Lett. 29, 19 (2002).
\bibitem{8}	T. Komabayashi, K. Hirose, N. Sata, Y. Ohishi, and L. S. Dubrovinsky, Phase Transition in CaSiO3 Perovskite, Earth Planet. Sci. Lett. 260, 564 (2007).
\bibitem{9}	Uchida T. et al., Non-cubic crystal symmetry of CaSiO3 perovskite up to 18 GPa and 1600 K, Earth Planet. Sci. Lett. 282, 268 (2009).
\bibitem{10}	L. Stixrude, R. E. Cohen, R. Yu, and H. Krakauer, Prediction of Phase Transition in CaSiO3 Perovskite and Implications for Lower Mantle Structure, Am. Mineral. 81, 1293 (1996).
\bibitem{11}	R. Caracas, R. Wentzcovitch, G. D. Price, and J. Brodholt, CaSiO3 Perovskite at Lower Mantle Pressures, Geophys. Res. Lett. 32, (2005).
\bibitem{12}	L. Li, D. J. Weidner, J. Brodholt, D. Alfè, G. D. Price, R. Caracas, and R. Wentzcovitch, Phase Stability of CaSiO3 Perovskite at High Pressure and Temperature: Insights from \textit{Ab Initio} Molecular Dynamics, Phys. Earth Planet. Inter. 155, 3–4 (2006).
\bibitem{13}	T. Sun, D.-B. Zhang, and R. M. Wentzcovitch, Dynamic Stabilization of Cubic CaSiO3 Perovskite at High Temperatures and Pressures from \textit{Ab Initio} Molecular Dynamics, Phys. Rev. B 89, 094109 (2014).
\bibitem{14}	J. C. A. Prentice, R. Maezono, and R. J. Needs, First-Principles Anharmonic Vibrational Study of the Structure of Calcium Silicate Perovskite under Lower Mantle Conditions, Phys. Rev. B 99, 064101 (2019).
\bibitem{15}	Z. Zhang and R. M. Wentzcovitch, \textit{Ab Initio} Anharmonic Thermodynamic Properties of Cubic Ca Si O 3 Perovskite, Phys. Rev. B 103, 10 (2021).
\bibitem{16}	A. R. Thomson, W. A. Crichton, J. P. Brodholt, I. G. Wood, N. C. Siersch, J. M. R. Muir, D. P. Dobson, and S. A. Hunt, Seismic Velocities of CaSiO3 Perovskite Can Explain LLSVPs in Earth’s Lower Mantle, Nature 572, 7771 (2019).
\bibitem{17}	T. Kurashina, K. Hirose, S. Ono, N. Sata, and Y. Ohishi, Phase Transition in Al-Bearing CaSiO3 Perovskite: Implications for Seismic Discontinuities in the Lower Mantle, Phys. Earth Planet. Inter. 145, 1–4 (2004).
\bibitem{18}	S. Ono, Y. Ohishi, and K. Mibe, Phase Transition of Ca-Perovskite and Stability of Al-Bearing Mg-Perovskite in the Lower Mantle, Am. Mineral. 89, 10 (2004).
\bibitem{19}	Murakami M., Hirose K., Sata N., and Ohishi Y., Post-perovskite phase transition and mineral chemistry in the pyrolitic lowermost mantle, Geophys. Res. Lett. 32, (2005).
\bibitem{20}	Adams D. J. and Oganov A. R., \textit{Ab Initio} molecular dynamics study of CaSiO3 perovskite at P-T conditions of Earth’s lower mantle, Phys. Rev. B 73, 184106 (2006).
\bibitem{21}	L. Stixrude, C. Lithgow-Bertelloni, B. Kiefer, and P. Fumagalli, Phase Stability and Shear Softening in CaSiO3 Perovskite at High Pressure, Phys. Rev. B 75, 2 (2007).
\bibitem{22}	D. Alfè, G. A. de Wijs, G. Kresse, and M. J. Gillan, Recent Developments in \textit{Ab Initio} Thermodynamics, Int. J. Quantum Chem. 77, 871 (2000).
\bibitem{23}	J. Behler and M. Parrinello, Generalized Neural-Network Representation of High-Dimensional Potential-Energy Surfaces, Phys Rev Lett 98, 14 (2007).
\bibitem{24}	A. P. Bartók, M. C. Payne, R. Kondor, and G. Csányi, Gaussian Approximation Potentials: The Accuracy of Quantum Mechanics, without the Electrons, Phys Rev Lett 104, 13 (2010).
\bibitem{25}	R. Jinnouchi, F. Karsai, and G. Kresse, On-the-Fly Machine Learning Force Field Generation: Application to Melting Points, Phys. Rev. B 100, 1 (2019).
\bibitem{26}	L. Zhang, J. Han, H. Wang, W. A. Saidi, R. Car, and W. E, End-to-End Symmetry Preserving Inter-Atomic Potential Energy Model for Finite and Extended Systems, arXiv:1805.09003.
\bibitem{27}	A. P. Thompson, L. P. Swiler, C. R. Trott, S. M. Foiles, and G. J. Tucker, Spectral Neighbor Analysis Method for Automated Generation of Quantum-Accurate Interatomic Potentials, J. Comput. Phys. 285, 316 (2015).
\bibitem{28}	T. D. Huan, R. Batra, J. Chapman, S. Krishnan, L. Chen, and R. Ramprasad, A Universal Strategy for the Creation of Machine Learning-Based Atomistic Force Fields, Npj Comput. Mater. 3, 1 (2017).
\bibitem{29}	J. Behler, Perspective: Machine Learning Potentials for Atomistic Simulations, J. Chem. Phys. 145, 170901 (2016).
\bibitem{30}	T. Morawietz, A. Singraber, C. Dellago, and J. Behler, How van Der Waals Interactions Determine the Unique Properties of Water, Proc. Natl. Acad. Sci. 113, 8368 (2016).
\bibitem{31}	M. Rupp, A. Tkatchenko, K.-R. Müller, and O. A. von Lilienfeld, Fast and Accurate Modeling of Molecular Atomization Energies with Machine Learning, Phys Rev Lett 108, 5 (2012).
\bibitem{32}	K. T. Schütt, F. Arbabzadah, S. Chmiela, K. R. Müller, and A. Tkatchenko, Quantum-Chemical Insights from Deep Tensor Neural Networks, Nat. Commun. 8, 1 (2017).
\bibitem{33}	S. Chmiela, A. Tkatchenko, H. E. Sauceda, I. Poltavsky, K. T. Schütt, and K.-R. Müller, Machine Learning of Accurate Energy-Conserving Molecular Force Fields, Sci. Adv. 3, e1603015 (2017).
\bibitem{34}	J. S. Smith, O. Isayev, and A. E. Roitberg, ANI-1: An Extensible Neural Network Potential with DFT Accuracy at Force Field Computational Cost, Chem. Sci. 8, 3192 (2017).
\bibitem{35}	F. Yang, Q. Zeng, B. Chen, D. Kang, S. Zhang, J. Wu, X. Yu, and J. Dai, Lattice Thermal Conductivity of MgSiO3 Perovskite and Post-Perovskite under Lower Mantle Conditions Calculated by Deep Potential Molecular Dynamics, Chin. Phys. Lett. 39, 116301 (2022).
\bibitem{36}	J. Deng, H. Niu, J. Hu, M. Chen, and L. Stixrude, Melting of MgSiO3 Determined by Machine Learning Potentials, Phys. Rev. B 107, 064103 (2023).
\bibitem{37}	T. Wan, C. Luo, Y. Sun, and R. M. Wentzcovitch, Thermoelastic Properties of Bridgmanite Using Deep Potential Molecular Dynamics, arXiv:2307.07127.
\bibitem{38}	H. Luo, B. B. Karki, D. B. Ghosh, and H. Bao, Anomalous Behavior of Viscosity and Electrical Conductivity of MgSiO3 Melt at Mantle Conditions, Geophys. Res. Lett. 48, e2021GL093573 (2021).
\bibitem{39}	L. Tang, C. Zhang, Y. Sun, K.-M. Ho, R. M. Wentzcovitch, and C.-Z. Wang, Structure and Dynamics of Fe90Si3O7 Liquids Close to Earth’s Liquid Core Conditions, Phys. Rev. B 108, 064104 (2023).
\bibitem{40}	Zhang C., Tang L., Sun Y., Ho K.-M., Wentzcovitch R. M., and Wang C.-Z., Deep machine learning potential for atomistic simulation of Fe-Si-O systems under Earth’s outer core conditions, Phys. Rev. Mater. 6, 063802 (2022).
\bibitem{41}	C. Luo, Y. Sun, and R. M. Wentzcovitch, High Throughput Sampling of Phase Space with Deep Learning Potentials: $\delta$-AlOOH at Geophysical Conditions, arXiv:2309.06712.
\bibitem{42}	X. Wang, Z. Wang, P. Gao, C. Zhang, J. Lv, H. Wang, H. Liu, Y. Wang, and Y. Ma, Data-Driven Prediction of Complex Crystal Structures of Dense Lithium, Nat. Commun. 14, 2924 (2023).
\bibitem{43}	T. Chen, F. Yuan, J. Liu, H. Geng, L. Zhang, H. Wang, and M. Chen, Modeling the High-Pressure Solid and Liquid Phases of Tin from Deep Potentials with \textit{Ab Initio} Accuracy, Phys. Rev. Mater. 7, 053603 (2023).
\bibitem{44}	L. Zhang, H. Wang, R. Car, and W. E, Phase Diagram of a Deep Potential Water Model, Phys. Rev. Lett. 126, 236001 (2021).
\bibitem{45}	G. Kresse and J. Hafner, \textit{Ab Initio} Molecular Dynamics for Liquid Metals, Phys. Rev. B 47, 558 (1993).
\bibitem{46}	G. Kresse and J. Furthmüller, Efficient Iterative Schemes for \textit{Ab Initio} Total-Energy Calculations Using a Plane-Wave Basis Set, Phys Rev B 54, 16 (1996).
\bibitem{47}	P. E. Blöchl, Projector Augmented-Wave Method, Phys Rev B 50, 24 (1994).
\bibitem{48}	J. P. Perdew, K. Burke, and M. Ernzerhof, Generalized Gradient Approximation Made Simple, Phys Rev Lett 77, 18 (1996).
\bibitem{49}	J. P. Perdew and A. Zunger, Self-Interaction Correction to Density-Functional Approximations for Many-Electron Systems, Phys. Rev. B 23, 5048 (1981).
\bibitem{50}	L. Zhang, J. Han, H. Wang, R. Car, and W. E, Deep Potential Molecular Dynamics: A Scalable Model with the Accuracy of Quantum Mechanics, Phys Rev Lett 120, 14 (2018).
\bibitem{51}	A. Togo and I. Tanaka, First Principles Phonon Calculations in Materials Science, Scr. Mater. 108, 1 (2015).
\bibitem{52}	Z. Zhang, D.-B. Zhang, T. Sun, and R. M. Wentzcovitch, Phq: A Fortran Code to Compute Phonon Quasiparticle Properties and Dispersions, Comput. Phys. Commun. 243, 110 (2019).
\bibitem{53}	D.-B. Zhang, T. Sun, and R. M. Wentzcovitch, Phonon Quasiparticles and Anharmonic Free Energy in Complex Systems, Phys. Rev. Lett. 112, 058501 (2014).
\bibitem{54}	R. Freitas, M. Asta, and M. De Koning, Nonequilibrium Free-Energy Calculation of Solids Using LAMMPS, Comput. Mater. Sci. 112, 333 (2016).
\bibitem{55}	K. Kawai and T. Tsuchiya, P-V-T Equation of State of Cubic CaSiO3 Perovskite from First-Principles Computation, J. Geophys. Res. Solid Earth 119, 2801 (2014).
\bibitem{56}	D. N. Sagatova, A. F. Shatskiy, N. E. Sagatov, and K. D. Litasov, Phase Relations in CaSiO3 System up to 100 GPa and 2500 K, Geochem. Int. 59, 8 (2021).
\bibitem{57}	N. Sun, Z. Mao, S. Yan, X. Wu, V. B. Prakapenka, and J.-F. Lin, Confirming a Pyrolitic Lower Mantle Using Self-Consistent Pressure Scales and New Constraints on CaSiO3 Perovskite, J. Geophys. Res. Solid Earth 121, 4876 (2016).
\bibitem{58}	A. Ricolleau, J.-P. Perrillat, G. Fiquet, I. Daniel, J. Matas, A. Addad, N. Menguy, H. Cardon, M. Mezouar, and N. Guignot, Phase Relations and Equation of State of a Natural MORB: Implications for the Density Profile of Subducted Oceanic Crust in the Earth’s Lower Mantle, J. Geophys. Res. 115, B08202 (2010).
\bibitem{59}	H. K. Mao, L. C. Chen, R. J. Hemley, A. P. Jephcoat, Y. Wu, and W. A. Bassett, Stability and Equation of State of CaSiO3-Perovskite to 134 GPa, J. Geophys. Res. 94, 17889 (1989).
\bibitem{60}	Y. Wang, D. J. Weidner, and F. Guyot, Thermal Equation of State of CaSiO3 Perovskite, J. Geophys. Res. Solid Earth 101, 661 (1996).
\bibitem{61}	H. Chen, S.-H. Shim, K. Leinenweber, V. Prakapenka, Y. Meng, and C. Prescher, Crystal Structure of CaSiO3 Perovskite at 28–62 GPa and 300 K under Quasi-Hydrostatic Stress Conditions, Am. Mineral. 103, 462 (2018).
\bibitem{62}	Y. Sun, M. I. Mendelev, F. Zhang, X. Liu, B. Da, C. Wang, R. M. Wentzcovitch, and K. Ho, \textit{Ab Initio} Melting Temperatures of Bcc and Hcp Iron Under the Earth’s Inner Core Condition, Geophys. Res. Lett. 50, e2022GL102447 (2023).
\bibitem{63}	T. Tadano, Y. Gohda, and S. Tsuneyuki, Anharmonic Force Constants Extracted from First-Principles Molecular Dynamics: Applications to Heat Transfer Simulations, J. Phys. Condens. Matter 26, 225402 (2014).
\bibitem{64}	T. Tadano and S. Tsuneyuki, Self-Consistent Phonon Calculations of Lattice Dynamical Properties in Cubic ${\mathrm{SrTiO}}_{3}$ with First-Principles Anharmonic Force Constants, Phys. Rev. B 92, 054301 (2015).
\bibitem{65}	K. Umemoto, K. Kawamura, K. Hirose, and R. M. Wentzcovitch, Post-Stishovite Transition in Hydrous Aluminous SiO2, Phys. Earth Planet. Inter. 255, 18 (2016).
\bibitem{66}	Z. Zhang, J. C. E. Irving, F. J. Simons, and T. Alkhalifah, Seismic Evidence for a 1000 Km Mantle Discontinuity under the Pacific, Nat. Commun. 14, 1714 (2023).
\bibitem{67}	B. Ko, E. Greenberg, V. Prakapenka, E. E. Alp, W. Bi, Y. Meng, D. Zhang, and S.-H. Shim, Calcium Dissolution in Bridgmanite in the Earth’s Deep Mantle, Nature 611, 7934 (2022).
	
	

\end{thebibliography}

\end{document}